# On operation and control of CW magnetrons for superconducting accelerators

G. Kazakevich[#], R.P. Johnson, Muons, Inc, Newport News, VA 23606, USA;
I. Gonin, V. Yakovlev, Fermilab, Batavia, IL 60510, USA;
Ya. Derbenev, H. Wang, JLab, Newport News, VA 23606, USA;

*Abstract*
CW magnetrons, developed for industrial RF heaters, were suggested to feed RF cavities of superconducting accelerators due to higher efficiency and lower cost of RF power than provide traditionally used klystrons, IOTs or solid-state amplifiers. RF amplifiers driven by a master oscillator serve as coherent RF sources. CW magnetrons are regenerative RF auto generators with a huge regenerative gain. This causes regenerative instability with a notable noise when a magnetron operates as an auto generator i.e., with the anode voltage above the threshold of self-excitation. Traditionally, an injection locking by a small signal is used for phase stabilization of magnetrons. In this case CW magnetrons with the injection-locked (coherent) oscillations generate a notable level of noise. This may preclude use of CW magnetrons in this mode in the Superconducting RF (SRF) accelerators. This paper reviews PIC modeling and previously obtained experimental results for the operation and control of CW magnetrons, which led to the development of techniques most suitable for various SRF accelerators using forced RF oscillation of magnetrons, when the magnetron is launched by an injected forcing signal, and regenerative noise is suppressed.

## 1. Introduction

High-power CW magnetrons, designed and optimized for industrial RF heaters, but driven by an injection-locking signal, were suggested in number of works to power superconducting RF (SRF) cavities in accelerators due to higher efficiency and lower cost of generated RF power per Watt than provide traditionally used RF amplifiers (klystrons, IOTs, solid-state amplifiers). The RF amplifiers driven by a master oscillator serve as coherent low noise RF sources. The CW magnetrons are regenerative RF auto generators with a huge regenerative gain of the resonant system to start up reliably with a self-excitation by noise even if the tube is powered by a DC power supply. Very large regenerative gain causes a regenerative instability with a notable regenerative noise.

The theory of magnetrons operation presently does not exist. The phenomenological models of magnetrons operation proposed 60-80 years ago generally relate to pulse magnetrons having much lower regenerative gain than CW tubes. Usually launching of pulse magnetrons provide high harmonics caused by short rise time of the modulating HV pulses.

---
[#] e-mail: gkazakevitch@yahoo.com
grigory@muonsinc.com

Thus, the old models cannot provide capabilities to choose the CW magnetrons operating parameters and designs for application of the tubes in various SRF accelerators.

Note, that simple techniques of injection-locking of a pulse 2 MW S-band magnetron by a wave (with the power of -18 dBc) reflected from the accelerating cavity of a microtron was implemented in the THz FEL project. The world's first terahertz FEL driven by the microtron, developed and built by BINP (Budker Institute of Nuclear Physics, Russia) for KAERI (Korea Atomic Energy Research Institute, South Korea) in 1999 was successfully put into operation. The techniques allowed an order of magnitude improvement in the intra-pulse stability of the magnetron, ensuring FEL operation in the 1-3 THz range with the peak power of up to 1 kW and the macro pulse power of ~50 W [1].

In [2] were reported first results of the r.m.s. phase deviations of ≈0.85 deg. obtained with a CW microwave oven magnetron driving an SRF 2.45 GHz cavity at 2 K. The measured r.m.s. value includes microphonics phase modulation with period of 30 Hz and modulation of the magnetron magnetic field with period of 60 Hz caused most likely by a filament AC power supply.

The general requirements for operation and control of CW magnetrons in SRF accelerators include phase and amplitude control to meet the accelerator performance providing the r.m.s. deviation in phase and amplitude of ≤0.5 deg. and 0.5% for Proton/ion accelerators and rings, of 0.1 deg. and 0.05%, for Nuclear physics accelerators, and 0.01 deg and 0.01% for Light sources [3].

Note that for now the microphonics r.m.s. deviations in phase and amplitude of 0.26 deg. and 0.3%, respectively, at 4 K in the same cavity were demonstrated in [4].

Typically, a CW magnetron is intended to operate in the self-excitation mode, with the anode voltage above the self-excitation threshold. A small injection-locking signal, $P_{Lock} \approx$ -20 dBc or less of the magnetron power $P_{Mag}$ were suggested to stabilize the tube. In this case a CW magnetron may start RF generation by the regenerative noise or ripples of Power Supplies (PS), which are uncorrelated with the injection-locking signal. We consider the probability of such RF generation vs. the injection-locking signal power $P_{Lock}$.

The experiments with 2.45 GHz microwave oven magnetrons [5] demonstrated that the standard requirement for the anode voltage ≥ the tubes self-excitation threshold is not the necessary condition for startup and RF generation of CW tubes. The experiments demonstrated startup and stable RF generation of CW microwave oven magnetrons fed notably below the self-excitation threshold voltage when they were driven by the injection-locking (forcing) signal ≥ -10 dBc.

There was developed and experimentally tested an analytical electrodynamic model for the launch and operation of CW magnetrons [6]. The model basing on the charge drift approximation [7] takes into account the resonant interaction of slow synchronous wave rotating in the magnetron interaction space with Larmor electrons moving toward the magnetron anode. The resonant interaction causes the resonant exchange of energy between moving charges and the synchronous wave rotating around the tube cathode. Resonant energy exchange occurs when the azimuthal velocities of the moving charge and the synchronous wave are approximately equal. In this case, the azimuthal electric field of the synchronous wave acts on the moving charges as a constant one providing the energy exchanges. If the azimuthal velocity of a moving charge is larger than azimuthal velocity of the rotating wave, the wave slows down the moving charge. This increment the energy into the wave and the magnetron resonant system resulting in phase grouping of the charges in "spokes". Otherwise, the wave accelerates the moving charge azimuthally, losing the energy in accordance with Landau theory on propagation of the waves in the electronic plasma [8]. This worsens the phase grouping of the drifting charges and may stop the magnetron RF generation. In study of the energy exchange in a magnetron were found conditions for the phase grouping of charges moving in the interaction space of the tube, necessary for its start-up and the most efficient and stable operation at the anode voltage below the self-excitation threshold.

The charge drift in a magnetron is determined by the superposition of the static electric field, described by the static electric potential $\Phi^0$, and the RF field of the synchronous wave, determined by a scalar potential $\Phi$, which is induced by the magnetron current and the injected resonant RF signal (in a steady state both are in phase). Thus, the drift of the charge can be described in the polar frame by the following system of equations of the first order [9]:

$$\begin{cases} \dot{r} = -\frac{c}{Hr}\frac{\partial}{\partial\varphi}(\Phi^0 + \Phi) \\ \dot{\varphi} = \frac{c}{Hr}\frac{\partial}{\partial r}(\Phi^0 + \Phi) \end{cases} \quad (1)$$

For the static electric field $\Phi^0 = U \ln(r/r_1)/\ln(r_2/r_1)$ and $E_r = \mathrm{grad}\ \Phi^0$, $d\Phi^0/d\varphi = 0$; therefore $E_\varphi(r) = 0$. Here $U$ is the magnetron anode voltage; $r_1$ and $r_2$ are the magnetron cathode and anode radii, respectively. The static radial electric field at the magnetron cathode is $E_r(r_1) = U/r_1 \ln(r_2/r_1)$. A slow RF (synchronous) wave type $\exp[-i(n\varphi + \omega t)]$ is excited at the frequency $\omega$ and is rotated in the space of interaction with the phase velocity $-\Omega = \omega/n$; $n = N/2$ ($N$ is the number of magnetron resonant cavities). This wave number sets the same azimuthal periodicity in interaction of the drifting charge with the RF field in the interaction space. The azimuthal velocity of the wave coincides with the azimuthal drift velocity of the center of the Larmor orbit located on the "synchronous" radius $r_S$: $r_S = \sqrt{-ncU/[\omega H \ln(r_2/r_1)]}$, ($H < 0$ is assumed). The scalar potential of the rotating wave $\Phi$ satisfying the Laplace equation with inaccuracy $\sim(2\pi r\Omega/c)^2$ is presented as:

$$\Phi = \sum_{k=-\infty}^{\infty} \frac{\tilde{E}_k(r_1)\cdot r_1}{2k}\left[\left(\frac{r}{r_1}\right)^k - \left(\frac{r_1}{r}\right)^k\right]\sin(k\varphi + \omega t), \quad (2)$$

Here $\tilde{E}_k(r_1)$ is the amplitude of the k-th harmonic of the radial RF electric field at $r = r_1$. The form of the potential was chosen so that the azimuthal electric field vanishes at the cathode. The coefficients $\tilde{E}_k(r_1)$ are determined by the requirement to have zeroed azimuthal electric field at the anode everywhere except the coupling slits of the cavities.

The term in the sum of Eq. (2) with $k = n$ has a resonant interaction with the azimuthal motion of the Larmor orbit. One considers only this term.

In the coordinate frame rotating with the synchronous wave for $\varphi_S = \varphi + t\cdot\omega/n$ and for an effective potential

$$\Phi_S = U\frac{\ln(r/r_1)}{\ln(r_2/r_1)} + \frac{\omega H}{2nc}r^2 + \frac{\tilde{E}_n(r_1)\cdot r_1}{2n}\times\left[\left(\frac{r}{r_1}\right)^n - \left(\frac{r_1}{r}\right)^n\right]\sin(n\varphi_S)$$

one obtains the drift equation system:

$$\begin{cases} \dot{r} = -\frac{c}{Hr}\frac{\partial}{\partial\varphi_S}\Phi_S \\ \dot{\varphi}_S = \frac{c}{Hr}\frac{\partial}{\partial r}\Phi_S \end{cases} \quad (3)$$

Substantiating $\Phi_S$ into Eqs. (3), and denoting: $\phi_0(r) = \ln\frac{r}{r_1} - \frac{1}{2}\left(\frac{r}{r_S}\right)^2$, and $\phi_1(r) = \frac{1}{2n}\left[\left(\frac{r}{r_1}\right)^n - \left(\frac{r_1}{r}\right)^n\right]$, one can obtain the system of drift equations in the frame of the synchronous wave expressed via the relative magnitude of resonant harmonic of the synchronous wave, $\varepsilon$, taken at cathode. The equations characterize the resonant interaction of the charge in center of Larmor orbit with the synchronous wave.

$$\begin{cases} \dot{r} = \omega\frac{r_S^2}{r}\varepsilon\phi_1(r)\cos(n\varphi_S) \\ n\dot{\varphi}_S = -\omega\frac{r_S^2}{r}\left(\frac{d\phi_0}{dr} + \varepsilon\frac{d\phi_1}{dr}\sin(n\varphi_S)\right) \end{cases} \quad (4)$$

Here: $\varepsilon = \tilde{E}_n(r_1)/E_r(r_1) = \tilde{E}_n(r_1)\cdot r_1 \ln(r_2/r_1)/U$.

Equations (4) show that without the synchronous wave $\varepsilon = 0$, $\dot{r} = 0$ and the drift of charges towards the tube anode is impossible.

The top equation describes the radial velocity of the moving charge. In accordance with this equation, the drift of the charge towards the anode is possible in the interval ($\pm\pi/2$) with the period of $2\pi$, i.e., only in "spokes". The charge can enter the spoke through the boundaries located at $n\varphi_S = \pm\pi/2$. The radial drift velocity is proportional to the synchronous wave magnitude, $\varepsilon$. The condition $\varepsilon \geq 1$ does not allow operation of the magnetron.

The second equation describes the azimuthal velocity of the drifting charge in the frame of the synchronous wave. The second term in the parentheses causes the phase grouping of the charge by the resonant RF field via potential $\phi_1$. The first term of this equation describes a radially-dependent azimuthal drift of the charge, resulting from the rotating frame, with azimuthal angular velocity $-\omega = \Omega/n$. This term at $r > r_S$ and at low $\varepsilon$ causes the movement of the charge from the phase interval allowed for spokes.

Equations (4) were integrated for a typical model of a commercial magnetron with $N = 8$, $r_1 = 5$ mm, $r_2/r_1 = 1.5$, $r_S/r_1 = 1.2$. Considering the charge drifting in the center of the Larmor orbit, were obtained the charge trajectories at $r \geq r_1 + r_L$ for various magnitudes $\varepsilon$ of the RF field in the synchronous wave and during time interval of the drift of $2 \leq \tau \leq 10$ cyclotron periods allowing coherent contribution to the synchronous wave.

The azimuthal boundaries of the charge drifting in a spoke at various $\varepsilon$ obtained by the integration are plotted in Fig. 1(a) assuming a uniform phase distribution of the emitted electrons. The phase interval at $(r_1+r_L)$ normalized by $\pi$ shows part of the emitted electrons (in the phase interval admitted for a spoke) contributing to the synchronous wave.

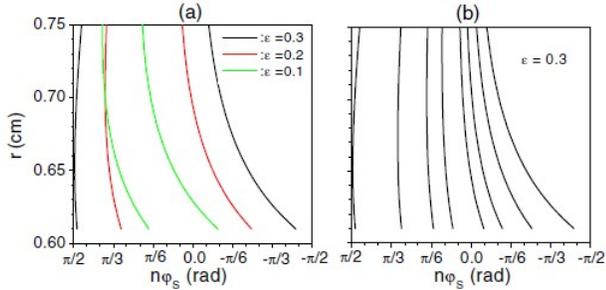

Fig. 1. Phase grouping of the charge drifting towards the magnetron anode in the considered magnetron model. Graph (a) shows the phase boundaries of trajectories of the charges contributing to the synchronous wave in dependence on $\varepsilon$. Graph (b) shows trajectories of the charges in a spoke at $\varepsilon = 0.3$.

A phase grouping of the drifting charges caused by resonant interaction with the rotating synchronous wave violates a phase uniformity of the initially emitted charges. This phenomenon may be explained as an appearance of degree of freedom of the drifting charges at the resonant interaction. Taking this into account and differentiating the bottom equation of system (4), one can obtain the impact of phase grouping on the back stream of electrons toward the cathode of the magnetron.

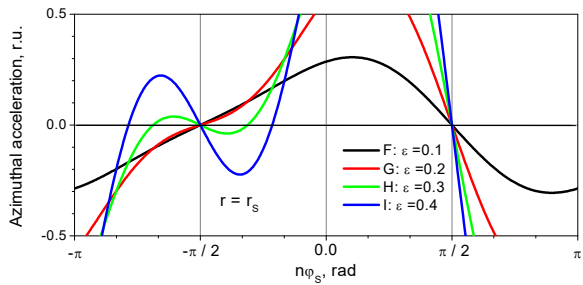

Fig. 2. An azimuthal acceleration shaping "spokes" of the drifting charges vs. $n\varphi_S$ at the radius of rotation of synchronous wave equal $r_S$ for various $\varepsilon$.

This plot one can interpret as an impact of a gradient of a grouping potential $\psi(r, n\varphi_S, \varepsilon)$ of the synchronous wave with a respective potential wells at the radius of rotation of the synchronous wave $r_S$.

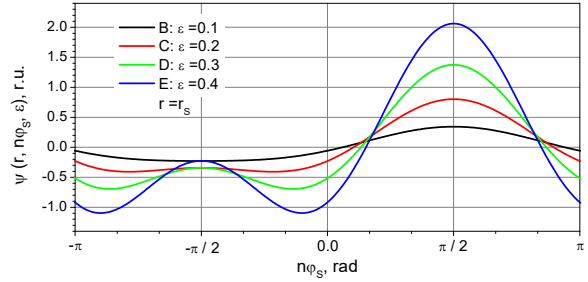

Fig. 3. A gradient of the grouping potential $\psi(r, n\varphi_S, \varepsilon)$ at the radius of rotation of synchronous wave $r_S$, vs. $\varepsilon$.

This graph shows losses of charges from the "spoke" at $\varepsilon < 0.3$ because of migration of part of charges from the phase interval ($\pm\pi/2$) allowed for a "spoke". This causes the electrons back stream in traditionally used regimes of magnetron operation. The emigrated electrons absorb the RF energy from the synchronous wave heating the cathode surface. Note, that in the developed electrodynamic model at $\varepsilon \geq 0.3$ (at the forcing signal of $\sim$ -10 dBc or more) the potential barriers prevent migration of charges from spokes for all radii of moving charges as it is presented below. This reduces overheat of the cathode surface by the quite strong phase grouping at the magnetrons operating with quite large forcing signal.

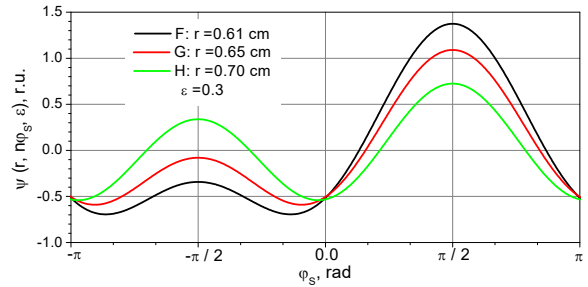

Fig. 4. Potential barriers preventing electron back stream at various radii of the drifting charges in a magnetron model at a large forcing signal.

Thus, taking into account the appearance of a degree of freedom considering the phase grouping of mowing charges in "spokes" rotating in the magnetron space of interaction indicates a notable reduction of the electron back stream that overheats the emitting surface when the tube operates with a quite large forcing signal in CW or pulse modes. PIC simulation performed for a microwave oven magnetron verified a decrease in the power of back-streaming electrons overheating the cathode in ~2 times.

The electrodynamic model substantiates unknown earlier, but most suitable for various SRF accelerators, methods of operation and control of CW magnetrons with a quite large (~ -10 dBc) injected locking (forcing) signal, allowing a wide range (≈10 dB) power control by varying the magnetron current [5]. The experiments showed that magnetron power with good accuracy linearly depends on the tube current; efficiency of the magnetron at such method of power control is highest, [ibid].

Measured V-I characteristic of CW, 2.45 GHz, 1.2 kW magnetron type 2M137-IL (from Richardson Electronics), Fig. 5, demonstrates the range of current control at -10 dBc forcing signal [6].

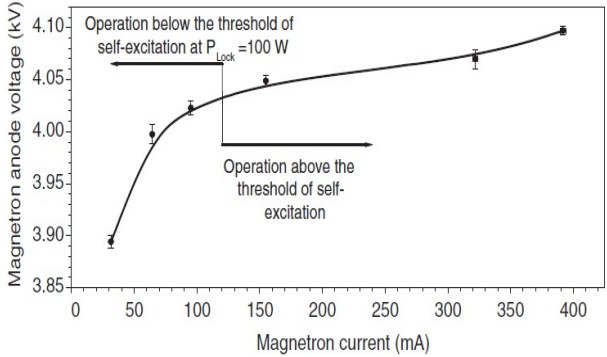

Fig. 5. Magnetron V-I characteristic measured at $P_{Lock}$ =100 W. The solid line (B-spline fit) shows the available range of current control with stable operation of the tube at power of the resonant injected signal $P_{Lock}$ =100 W.

## 2. On stable operation of a CW magnetron below the self-excitation threshold voltage

The carrier frequency offset of CW magnetron type 2M137-IL stably operating at the output power of 100 W was measured vs. $P_{Lock}$ in experiments at RBW =100 kHz and the anode voltage by ≈150 V less than the self-excitation threshold of 4.04 kV. (Ref -45 dBc) [5]. The offset is shown in the Fig. 6.

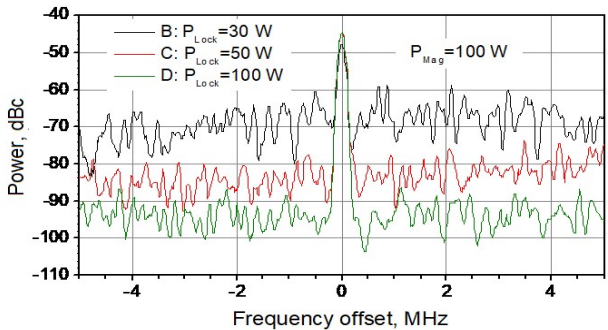

Fig. 6. Offset of the carrier frequency of the magnetron type 2M137-IL at the output power of 100 W, measured with RBW =100 kHz at the anode voltage by ≈150 V less than the self-excitation threshold, Ref. -45 dBc vs. $P_{Lock}$ [5].

One can see presence of the forcing signal peaks in all traces at the output since the signal is applied to the tube output. The shape of oscillations eliminating the injection-locking signal from the output power offset one can get by method of adjacent averaging of the traces. This method may significantly reduce sharp peaks keeping smooth spectra, Fig. 7.

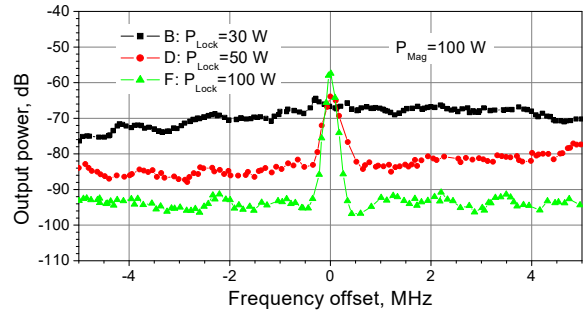

Fig. 7. Smoothed offset of the carrier frequency of the 2M137-IL magnetron at the magnetron output power of 100 W, Ref. -45 dB, vs. power of the injection-locking signal, $P_{Lock}$ [10].

In this graph the trace with $P_{Lock}$ =30 W shows the broad-band RF oscillations in the interaction space at the bandwidth of ≈7 MHz (that is determined by the tube resonant system bandwidth) at the level of -3 dB amplified by a resonant system up to output power of ≈1 W. One can infer that this trace shows the amplified incoherent spontaneous RF oscillations close to point of launching i.e., in fact, the quasi-continuous noise. Other traces show significantly reduced incoherent spontaneous oscillations due to conversion into the coherent one when the magnetron is launched [10]. At $P_{Lock}$ =50 W the spontaneous oscillations become partially coherent (partially phase-locked) due to larger azimuthal electric field of the synchronous wave causing the phase grouping. The residual spontaneous oscillations power characterizes the loss of coherency vs. $P_{Lock}$ at the phase grouping. A further increase of the locking (forcing) signal to 100 W additionally reduces power of the spontaneous oscillations converting them mainly into coherent one.

A noticeable frequency shift of the incoherent spontaneous oscillations at $P_{Lock}$ =50 W may indicate the frequency pushing caused by the anode current loading the resonant system of launched magnetron.

Thus in CW microwave oven magnetrons, the total phase-locking can be realized at the corresponding power of the locking (forcing) signal ≥ -10 dBc.

Further, using the electrodynamic model, the techniques of forced RF generation of CW magnetrons eliminating the launching of magnetrons by regenerative noise and PS ripples was developed and described. The techniques was verified in experiments with CW magnetrons for microwave ovens.

The performed PIC simulation [11] of type 2M253, 2.45 GHz, 0.86 kW, CW magnetron with nominal anode voltage of 4.0 kV shows RF oscillations in the interaction space at the magnetic field of 0.18 T with the locking (forcing) signal of -8 dBc is plotted in Fig. 8, upper graph.

This graph shows Landau damping of the injected resonant signal during of about 50 ns and slow buildup of superposition of the oscillation caused by currents in

"spokes" (at the frequency $-\Omega=\omega/n$) and the phase-locked oscillation at the frequency $\omega$.

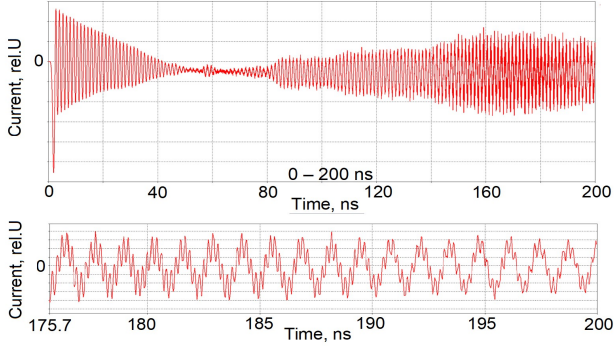

Fig. 8. Simulated superposition of oscillations of currents in the interaction space at the magnetron anode voltage of 4.10 kV and the injected resonant signal of 170 W.

The lower graph with delay of 175 ns shows the oscillation by currents in "spokes" $(-\Omega = \omega/n)$ and the phase-locked oscillation at the frequency $\omega$.

An increase of the forcing signal Fig. 9 reduces buildup time of the phase-locked oscillations.

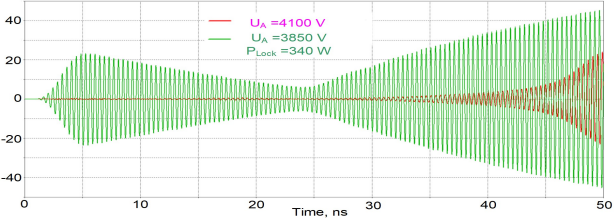

Fig. 9. Red oscillations show slow buildup of the magnetron free run operation at the anode voltage 4.10 kV. Green oscillations show buildup of the magnetron driven by a large forcing signal (of about -5 dBc) with the anode voltage below the self-excitation threshold. At this anode voltage the magnetron cannot startup the auto generation.

It is clearly seen that 2.45 GHz oscillation at the magnetic field of ≈0.18 T with the cyclotron second harmonic's period of ≈0.2 ns, rising during of about 10 periods (that is equal to period of the slow synchronous rotating wave) damps in about 20 ns before notable grows caused by filling of the resonant structure. The Landau damping process [8] is caused by insufficient phase grouping of Larmor electrons until the azimuthal velocities of drifting charges reach the values that provide increments of the synchronous wave and the RF field in a resonant system with a Q-factor of ~100. A smaller locking signal increases the start time of the phase-locked oscillations reducing the probability of injection locking [12]. Thus, even at a quite large locking signal a CW magnetron may generate the incoherent noise.

## 2. Operation of a CW injection-locked magnetron in the self-excitation mode

We consider operation of a CW magnetron as it is traditionally assumed, in the self-excitation mode, at an injection-locking signal with power $P_{Lock}$.

The effective bandwidth of injection-locking $\Delta f$, at the locking signal $P_{Lock}$ is expressed by the following equation [12]:

$$\Delta f = \frac{f_0}{2Q_L}\sqrt{\frac{P_{Lock}}{P_{Mag}}}. \quad (5)$$

Here: $f_0$ is, in fact, the magnetron instantaneous frequency, $Q_L$ is the magnetron loaded Q-factor; $P_{Mag}$ is the magnetron output power. For free running CW, 2.45 MHz magnetrons the spectra width is of about 2-4 MHz at -3 dBc. The spectra consist of numerous startups with the instantaneous frequencies filling the spectral line width [13]. Out of the effective bandwidth the magnetron cannot be injection-locked at the given $P_{Lock}$ as it follows from Adler's theory [14].

Then the probabilities of the injection-locking process $w_{Lock}$ and a free running operation $w_{FR}$ ($P_{Lock} =0$) for the 2.45 GHz CW tube vs. $P_{Lock}$ one can estimate as [10]:

$$w_{Lock} \sim \frac{\Delta f}{\Delta f_{FR}}. \quad (6)$$

$$w_{FR} \sim \frac{\Delta f_{FR} - \Delta f}{\Delta f_{FR}}. \quad (7)$$

The estimates of $\Delta f$, $w_{Lock}$ and $w_{FR}$ vs. $P_{Lock}$ are shown in Table 1.

Table 1. The estimates of $\Delta f$, $w_{Lock}$ and $w_{FR}$ vs. $P_{Lock}$.

| $P_{Lock}$ | $\Delta f$, | $w_{Lock}$ | $w_{FR}$ |
|---|---|---|---|
| -10 dB | 3.87 MHz | ~0.97 | ~0.03 |
| -20 dB | 1.22 MHz | ~0.3 | ~0.7 |
| -30 dB | 0.39 MHz | ~0.1 | ~0.9 |

Thus, the probability of the injection-locked RF generation of such RF source at low injection-locking signal may be notably less than probability of the free running generation caused by noise. If the noise oscillations are incoherent, they are much less in magnitude than the injection-locked (coherent) ones. These incoherent oscillations result in a notable quasi-continuous broadband noise spectrum which is a disadvantage for operation of the auto generating magnetrons with low locking signal for high Q-factor SRF cavities. Note, if the effective bandwidth of the injection-locking is approaching zero, i.e., the transient time of the injection locking tends to infinity [12], then probability of the injection-locking process is approaching zero as well.

Impact of the injection-locking signal power on the noise was studied measuring the spectral density of noise power [6] vs. locking power, $P_{Lock}$ with 2.45 GHz magnetron type 2M137-IL at $P_{Lock}$ =10 W, 30 W, and 100 W respectively, (traces C, B, A) at the magnetron output power of 1 kW in CW mode, Fig. 10.

This figure shows that the spectral density of noise power of the generating tube (interval 0.1 MHz-1.0 MHz) is increased by 20 dBc/Hz at $P_{Lock}$ =10 W (-20 dBc), comparing to measurements at $P_{Lock}$ =100 W (-10 dBc). This indicates low probability of the phase locking at the locking signal of 10 W.

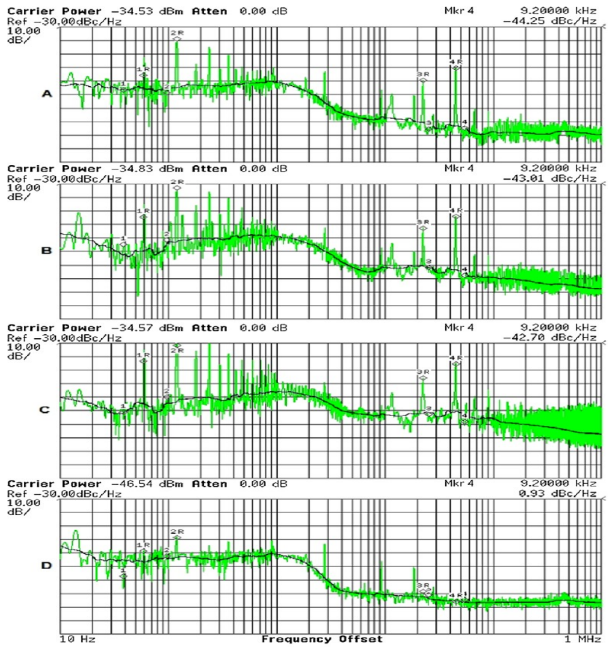

Fig. 10. The spectral density of noise power of the generating magnetron (in the interval 0.1 MHz-1.0 MHz). Traces A at $P_{Lock}$ =100 W, traces B at $P_{Lock}$=30W, traces C at $P_{Lock}$ =10 W. Trace D is the spectral density of the locking signal noise power, when the tube anode voltage was OFF.

## 3. Operation of CW magnetrons above and below the self-excitation threshold voltage

Measured in CW mode at the bandwidth resolution of 5 Hz the carrier frequency offsets Fig. 11, show peaks and quasi-continuous noise spectra of the magnetron type 2M137-IL when the tube operates above or below the self-excitation threshold voltage at various injected resonant signals $P_{Lock}$ and the output RF powers $P_{Mag}$ [5].

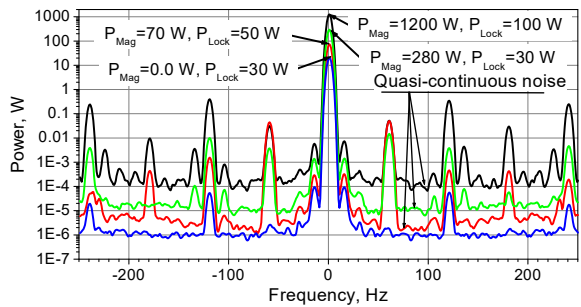

Fig. 11. Offsets of the carrier frequency of the magnetron type 2M137-IL with the threshold of self-excitation of 4.04 kV measured in CW mode at the various RF powers $P_{Mag}$, and the injection-locking (forcing) signal, $P_{Lock}$.

The trace $P_{Mag}$ =0.0 W, $P_{Lock}$ =30 W is the carrier frequency offset of the injection-locking signal when the magnetron anode voltage was OFF. The traces $P_{Mag}$ =70 W, $P_{Lock}$ =50 W and $P_{Mag}$ =280 W, $P_{Lock}$ =30 W show the tube operation below the self-excitation threshold voltage. The trace $P_{Mag}$ =1200 W, $P_{Lock}$ =100 W is related to operation above the self-excitation voltage. The tube anode voltages $U_{Mag}$ in these measurements were 3.90, 4.01 and 4.09 kV, respectively. Note the very narrow width of the measured spectral lines of the magnetron, which is most suitable for powering high-Q SRF cavities.

A noticeable level of quasi-continuous noise of magnetrons operating above the self-excitation threshold voltage indicates a high probability of incoherent generation, i.e., a reduced probability of the injection locking process. For accelerators, this may cause problems with suppressing parasitic modulations of the accelerating field in SRF cavities.

The ratios of the carrier frequency peak power to the quasi-continuous noise power (excluding sidebands caused by PS ripples) reach a maximum when the magnetron operates below the self-excitation threshold voltage. This makes operation of CW tubes below the self-excitation threshold more suitable for superconducting RF accelerators.

## 4. Stimulated RF generation mode of a CW magnetron

A novel techniques realizing the Stimulated RF generation mode was experimentally verified in a pulse regime using the CW 2.45 GHz magnetron type 2M219G (nominal output power of 945 W and the measured self-excitation threshold voltage of 3.69 kV) [15]. The mode uses quite large injected forcing signal and the anode voltage is chosen to eliminate any launchings of the tube by noise or PS ripples, i.e., eliminating auto generation, reducing the regenerative gain. The pulse regime with train of pulses showing absence of magnetron startups caused by noise or PS ripples in intervals between the train pulses was used.

The forcing signal excites a rotating synchronous wave in the interaction space with an azimuthal electric field sufficient for phase locking of the spontaneous oscillations and improves (increases) phase grouping of the charges mowing in the interaction space. This, in turn, increases the azimuthal electric field of the synchronous wave [6]. The incoherent spontaneous emission by the phase locking is transformed into a coherent oscillation amplified by a resonant system of a magnetron. Setup to study the Stimulated RF generation mode in pulse regime by a gated forcing signal is shown in Fig. 12 [15]. Main characteristics of RF generation in this mode are presented below.

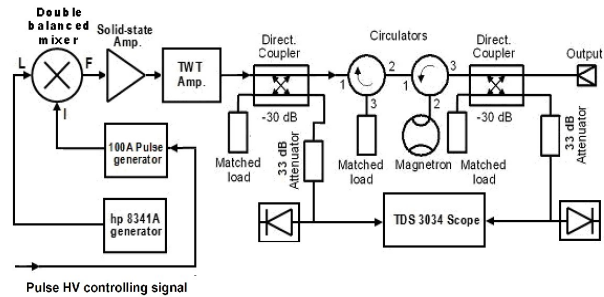

Fig. 12. Setup to study pulse control of the magnetron

type 2M219G driven by HP 8341A generator via the double balance mixer ZEM-4300MH (from Mini-Circuits) controlled by the pulse generator type 100A, and via solid state and TWT amplifiers.

The solid lines in Fig. 13 characterize the Stimulated RF generation mode showing the measured ranges of the anode voltage and magnetron current depending on the power of the injected forcing signal $P_{Lock}$. The magnetron was powered by a pulse High Voltage (HV) source [16], using a partial discharge of the 200 µF storage capacitor for pulse anode voltage. The pulse HV source was powered by a charging Glassman 10 kV, 100 mA switching power supply allowing for voltage control. Measured ranges of the magnetron RF generation were obtained at the forcing RF signal with duration of 2.2 ms at the pulse anode voltage with duration of 5.1 ms [15]. The magnetron current was measured by the current transducer type LA 55-P at the integration time of the measuring circuit of about 50 µs.

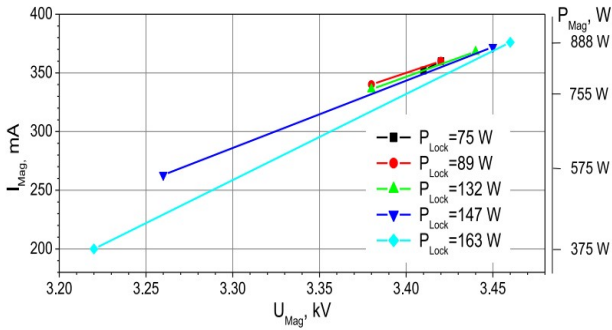

Fig. 13. Ranges of the anode voltage and current of the 2M219G magnetron operating in the Stimulated RF generation mode at different power levels of the injected resonant forcing signal $P_{Lock}$. The right scale shows measured RF output power $P_{Mag}$ of the magnetron corresponding to the magnetron current, $I_{Mag}$ [15].

As it follows from Fig. 13, the injected resonant signal of 163 W (-7.6 dBc) provides the range of the magnetron power control of ≈3.7 dB by variation of the magnetron current. An increase of the range of the tube power (current) control is caused by larger coherent gain [17] due to improved phase grouping at larger injected resonant forcing signal.

The magnetron performs pulse Stimulated RF generation by injection of a pulse forcing resonant signal into the tube during a DC feeding power Figs. 14, 15. The shapes and values of the forcing and output RF signals were measured by detectors with zero-bias Schottky diodes calibrated to better than ±0.5%.

Despite the significantly lower regenerative gain at a reduced anode voltage, the magnetron in the Stimulated RF generation mode with a large forcing signal converts the spontaneous oscillations in the interaction space into coherent ones [10], which are amplified by the resonant system of the tube. This ensures virtually nominal output RF power with high coherent gain that compensates reduced regenerative gain [17].

The absence of noise at the absence of a forcing resonant injected signal in the Stimulated RF generation mode is represented by the traces in Figs. 14 and 15 [15].

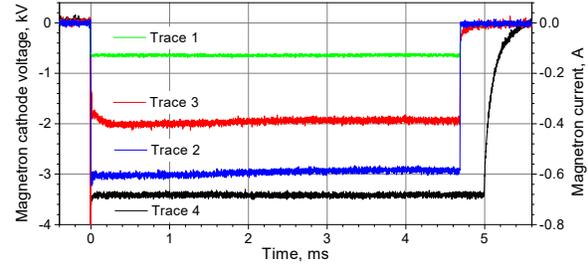

Fig. 14. Pulse operation of the magnetron in the stimulated RF generation mode with pulse duration of 4.7 ms. Traces 1 and 2 are the resonant injected and the magnetron output RF signals with powers of 74 W and 888 W, respectively; trace 3 is the magnetron pulsed current measured by the current transducer (right scale); trace 4 is the magnetron cathode voltage (−3.43 kV).

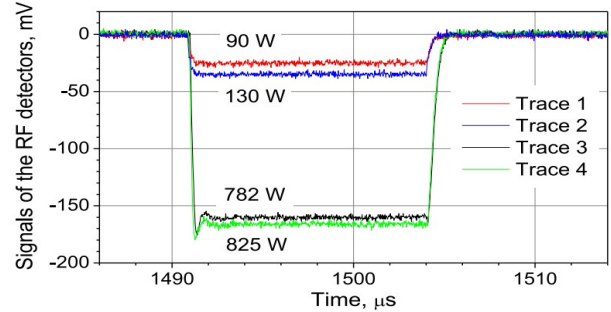

Fig. 15. Measured pulses of the 13 µs 20 kHz train when the magnetron operates in the Stimulated RF generation mode. Traces 3 and 4 are the magnetron output RF signals in dependence on the power of forcing signals, traces 1 and 2, respectively. The measurements were performed at the magnetron anode voltage of 3.41±0.01 kV.

The conversion efficiency $\eta$, expressed by the ratio of the generated RF power $P_{RF}$ to the consumed power of a CW magnetron operating in the Stimulated RF generation mode, neglecting the filament power is shown below.

$$\eta \approx P_{RF}/(U_{Mag} \cdot I_{Mag} + P_{Lock}) \qquad (8)$$

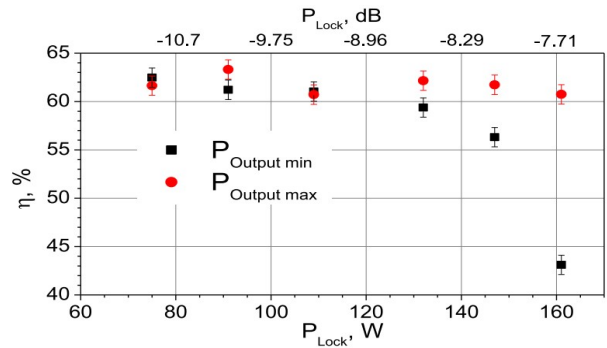

Fig. 16. Dependence of conversion efficiency of the 2M219G magnetron on power of injected forcing signal $P_{Lock}$ [15].

The plotted measured results are related to the maximum and minimum powers of the magnetron operating in the Stimulated RF generation mode. This demonstrates the range of the magnetron power (current) control vs. $P_{Lock}$. This method of power control in a wide range is the most efficient. The measured conversion efficiency of the magnetron operating above the self-excitation threshold in the "free run" mode ($P_{Lock} = 0$) at the nominal tube power is ≈54%.

Usage of two-cascade magnetrons allows reducing the necessary power of the forcing signal for the Stimulated RF generation mode down to ≈ -20 dBc [6]. In this scheme both magnetrons operate in the Stimulated RF generation mode; the first one with power ≈10% of the required maximum RF power drives the second high-power tube realizing the required power control by variation of the current in the high-power magnetron. The cost of two-cascade magnetrons in mass production should be significantly lower than the cost of traditionally used RF sources.

## 5. Bandwidth of phase and power control of magnetrons in Stimulated RF generation mode

For various SRF accelerator projects (colliders, ADS facilities and even industrial SRF accelerators) is important wide-band control of SRF sources in phase and power for suppression various parasitic modulations (microphonics, etc. The Stimulated RF generation mode utilizes a sufficiently large resonant forcing signal, which is also necessary for wide-band control; it is most suitable for magnetron RF sources for various SRF accelerators.

The bandwidth of control $BW_C$ for 2.45 GHz microwave oven magnetrons was obtained measuring the transfer functions magnitude and phase characteristics [6]. This is presented in Fig. 17 [10].

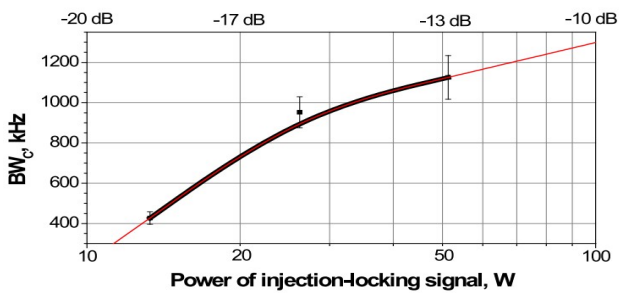

Fig. 17. The admissible bandwidth of control of 2.45 GHz microwave oven magnetrons determined by measured transfer functions characteristics. Black bold line shows the range and results of measurements with B-spline fit; the thin red line shows extrapolation.

The plotted curves indicate that even for 650 MHz magnetron RF sources the bandwidth of control should be ~1 MHz when they operate in the Stimulated RF generation mode.

Estimates of the bandwidths of phase and power control necessary for suppression of microphonics in ADS facility with 1 GeV, SRF, 650 MHz proton driver for the proton beam from 1 to 10 mA are presented in [17].

## 6. Summary

Measurements of the spectral density of noise power of CW microwave oven magnetrons locked by an injected signal ≤ -20 dBc indicate that the effective bandwidth of injection-locking is narrow as it is predicted by Adler theory. It means low probability of the injection-locked RF generation of the magnetron. In this case with quite high probability the magnetron generates noise. Increasing the injection-locking signal to -10 dBc reduces the noise spectral power density by ~20 dBc/Hz. This indicates significantly increased probability of the injection-locking process at $P_{Lock}$ = -10 dBc.

Measurements show that the quasi-continuous noise spectra are significantly reduced in operation of the CW magnetrons below the threshold of self-excitation.

The developed Stimulated RF generation mode eliminates the quasi-continuous noise of CW magnetrons at the absence of the injected forcing signal. It means that magnetron in this mode generates the oscillations forced by the injected resonant signal, i.e., the forced coherent oscillations like traditional RF amplifiers.

Efficiency of CW magnetrons in the Stimulated RF generation mode is higher than efficiency of magnetrons operating above the self-excitation threshold voltage: in a free run mode or driven by a small injection-locking signal, see [5].

The Stimulated RF generation mode allows for control of the magnetron power with highest efficiency, changing the magnetron current in a wide range, up to almost the nominal power of the tube.

The Stimulated RF generation mode is suitable for CW and pulse SRF accelerators. However, when using the Stimulated RF generation mode for pulsed SRF accelerators, high-voltage pulse modulators shaping the pulse anode voltage of the magnetrons are not required. The Stimulated RF generation mode provides 100% pulse modulation of the magnetron output power at 100% pulse modulation of the injected forcing signal.

The bandwidth of the phase and power control in CW magnetrons operating in the Stimulated RF generation mode is most suitable for suppression of parasitic modulations in the SRF cavities of accelerators.

Reduced injected forcing signal necessary for the Stimulated RF generation mode provide two-cascade magnetrons. This allows the reduction of power of the injected resonant signal by ≈10 dB. The first CW magnetron with $P_{Mag}$ ≈10% of the power required from the RF source provides the injected forcing signal for the second, high power CW magnetron. The power control is realized by regulation of the magnetron current in the high-power tube. At the mass production the cost of the cascaded RF source will be increased insignificantly.

The forcing signal with quite high power used for Stimulated RF generation mode notably reduces the electron back-stream overheating the magnetron cathode; this is important to improve the magnetron longevity.

## Acknowledgment

This manuscript has been authored by collaboration of Muons, Inc - Fermilab under CRADA-FRA-2017-0026 and CRADA FRA 2023-0029 between Muons, Inc., and Fermilab under FermiForward Discovery Group, LLC, Operator of Fermi National Accelerator Laboratory under contract No. 89243024CSC000002 with the U.S. Department of Energy, Office of Science, Office of High Energy Physics.